\documentclass[11pt]{article}\textwidth 6.5in\textheight 9in
\usepackage{amssymb}\usepackage[colorlinks]{hyperref}\usepackage{color}
	\usepackage{stmaryrd}\usepackage{mathrsfs}
	\usepackage{graphicx}
	\usepackage{amsmath}
\usepackage{caption}

\begin{document}
\setlength{\captionmargin}{27pt}
\newcommand\hreff[1]{\href {http://#1} {\small http://#1}}
\newcommand\trm[1]{{\bf\em #1}} \newcommand\emm[1]{{\ensuremath{#1}}}
\newcommand\prf{\paragraph{Proof.}}\newcommand\qed{\hfill\emm\blacksquare}

\newtheorem{thr}{Theorem} 
\newtheorem{lmm}{Lemma}
\newtheorem{cor}{Corollary}
\newtheorem{con}{Conjecture} 
\newtheorem{prp}{Proposition}

\newtheorem{blk}{Block}
\newtheorem{dff}{Definition}
\newtheorem{asm}{Assumption}
\newtheorem{rmk}{Remark}
\newtheorem{clm}{Claim}
\newtheorem{example}{Example}

\newcommand{\ab}{a\!b}
\newcommand{\yx}{y\!x}
\newcommand{\yux}{y\!\underline{x}}

\newcommand\floor[1]{{\lfloor#1\rfloor}}\newcommand\ceil[1]{{\lceil#1\rceil}}

\newcommand{\lea}{<^+}
\newcommand{\gea}{>^+}
\newcommand{\eqa}{=^+}

\newcommand{\lel}{<^{\log}}
\newcommand{\gel}{>^{\log}}
\newcommand{\eql}{=^{\log}}

\newcommand{\lem}{\stackrel{\ast}{<}}
\newcommand{\gem}{\stackrel{\ast}{>}}
\newcommand{\eqm}{\stackrel{\ast}{=}}

\newcommand\edf{{\,\stackrel{\mbox{\tiny def}}=\,}}
\newcommand\edl{{\,\stackrel{\mbox{\tiny def}}\leq\,}}
\newcommand\then{\Rightarrow}

\newcommand\km{{\mathbf {km}}}\renewcommand\t{{\mathbf {t}}}
\newcommand\KM{{\mathbf {KM}}}\newcommand\m{{\mathbf {m}}}
\newcommand\md{{\mathbf {m}_{\mathbf{d}}}}\newcommand\mT{{\mathbf {m}_{\mathbf{T}}}}
\newcommand\K{{\mathbf K}} \newcommand\I{{\mathbf I}}

\newcommand\II{\hat{\mathbf I}}
\newcommand\Kd{{\mathbf{Kd}}} \newcommand\KT{{\mathbf{KT}}} 
\renewcommand\d{{\mathbf d}} 
\newcommand\D{{\mathbf D}}

\newcommand\w{{\mathbf w}}
\newcommand\Ks{\mathbf{Ks}} \newcommand\q{{\mathbf q}}
\newcommand\E{{\mathbf E}} \newcommand\St{{\mathbf S}}
\newcommand\M{{\mathbf M}}\newcommand\Q{{\mathbf Q}}
\newcommand\ch{{\mathcal H}} \renewcommand\l{\tau}
\newcommand\tb{{\mathbf t}} \renewcommand\L{{\mathbf L}}
\newcommand\bb{{\mathbf {bb}}}\newcommand\Km{{\mathbf {Km}}}
\renewcommand\q{{\mathbf q}}\newcommand\J{{\mathbf J}}
\newcommand\z{\mathbf{z}}

\newcommand\B{\mathbf{bb}}\newcommand\f{\mathbf{f}}
\newcommand\hd{\mathbf{0'}} \newcommand\T{{\mathbf T}}
\newcommand\R{\mathbb{R}}\renewcommand\Q{\mathbb{Q}}
\newcommand\N{\mathbb{N}}\newcommand\BT{\{0,1\}}
\newcommand\FS{\BT^*}\newcommand\IS{\BT^\infty}
\newcommand\FIS{\BT^{*\infty}}\newcommand\C{\mathcal{L}}
\renewcommand\S{\mathcal{C}}\newcommand\ST{\mathcal{S}}
\newcommand\UM{\nu_0}\newcommand\EN{\mathcal{W}}

\newcommand{\supp}{\mathrm{Supp}}

\newcommand\lenum{\lbrack\!\lbrack}
\newcommand\renum{\rbrack\!\rbrack}
\renewcommand\i{\mathbf{i}}
\renewcommand\qed{\hfill\emm\square}

\title{Regression and Algorithmic Information Theory}

\author {Samuel Epstein\footnote{JP Theory Group. samepst@jptheorygroup.org}}

\maketitle

\renewcommand{\contentsname}{\centering Contents}
\begin{abstract}
In this paper we prove a theorem about regression, in that the shortest description of a function consistent with a finite sample of data is less than the combined conditional Kolmogorov complexities over the data in the sample.
\end{abstract}

\section{Introduction}
Classification is the task of learning a binary function $c$  from $\N$ to bits $\BT$. The learner is given a sample consisting of pairs $(x,b)$ for string $x$ and bit $b$ and outputs a binary classifier $h:\N\rightarrow\BT$ that should match $c$ as much as possible. Occam's razor says that ``the simplest explanation is usually the best one.'' Simple hypothesis are resilient against overfitting to the sample data. With certain probabilistic assumptions, learning algorithms that produce hypotheses of low Kolmogorov complexity are likely to correctly predict the target function~\cite{BlumerEhHaWar89}. The following theorem \cite{Epstein21} shows that the samples can be compressed to their count.\\

\noindent\textbf{Theorem.} \textit{Given a set of samples $\{(x_i,b_i)\}_{i=1}^n$, there is a function $f:\N\rightarrow\BT$ such that $f(x_i)=b_i$, for $i=1,\dots,n$, and $\K(f)\lel n + \I(\{(x_i,b_i)\};\ch)$.}\\

The $\K$ term is Kolmogorov complexity and the $\I$ term is defined in Section \ref{sec:conv}. Another area of machine learning is regression, in which one is given a set of pairs $\{(x_i,y_i)\}$, $i=1\dots n$, and the goal is to find a function $f$, such that $f(x_i)=y_i$. Usually each $x_i$ and $y_i$ represents a point in Euclidean space, but for our purposes they are natural numbers. As in classification, the goal is to use Occam's razor to find the simpliest function, to prevent overfitting to the random noise inherent in the sample data. This paper presents the following bounds on the simplest total computable function completely consistent with the data.\\

\noindent\textbf{Theorem.} \textit{For $\{(x_i,y_i)\}_{i=1}^n$,  there exists $f:\N\rightarrow\N$ with $f(x_i)= y_i$ for $i\in\{1,\dots,n\}$ and $\K(f) \lel \sum_{i=1}^n\K(y_i|x_i)+\I(\{(x_i,y_i)\};\ch)$.}
\section{Conventions}
\label{sec:conv}
For positive real functions $f$, by ${\lea}f$, ${\gea}f$, ${\eqa}f$, and ${\lel} f$, ${\gel} f$, ${\sim} f$ we denote ${\leq}\,f{+}O(1)$, ${\geq}\,f{-}O(1)$, ${=}\,f{\pm}O(1)$ and ${\leq}\,f{+}O(\log(f{+}1))$, ${\geq}f\,{-}O(\log(f{+}1))$, ${=}\,f{\pm}O(\log(f{+}1))$. $\K(x|y)$ is the conditional prefix Kolmogorov complexity. The chain rule states $\K(x,y)\eqa \K(x) +\K(y|\K(x),x)$.  Let $[A]=1$ if the mathematical statement $A$ is true, otherwise $[A]=0$. Let $\K_t(x|y)=\inf\{\|p\|:U_y(p)=x \textrm{ in $t$ steps}\}$. The information the halting sequence $\ch$ has about $x$ is $\I(x;\ch|y)=\K(x|y)-\K(x|y,\ch)$. $\I(x;\ch)=\I(x;\ch|\emptyset)$. A probability measure is elementary if its support is finite and it has rational values. The deficiency of randomness of $x\in\FS$ with respect to elementary probability measure $Q$ is $\d(X|Q)=\ceil{-\log Q(X)-\K(x|\langle Q\rangle)}$. The stochasticity of $x$ is $\Ks(x)=\min_Q \K(Q)+3\log\max\{\d(X|Q),1\}$.
\begin{lmm}[\cite{Epstein21,Levin16}]
\label{lmm:ksh}
$\Ks(x)\lel\I(x;\ch)$.
\end{lmm}

\begin{lmm}[\cite{EpsteinDerandom22}]
\label{lmm:cons}For partial computable $f$, $\I(f(x):\ch)\lea \I(x;\ch)+\K(f)$.
\end{lmm}

\section{Results}

Let $\Omega = \sum\{2^{-\|p\|}:U(p)\textrm{ halts}\}$ be Chaitin's Omega, $\Omega_n\in\Q_{\geq 0}$ be be the rational formed from the first $n$ bits of $\Omega$, and $\Omega^t = \sum\{2^{-\|p\|}:U(p)\textrm{ halts in time $t$}\}$. For $n\in \N$, let $\bb(n) = \min \{ t : \Omega_n<\Omega^t\}$. $\bb^{-1}(m) = \arg\min_n \{\bb(n-1)<m\leq \bb(n)\}$. Let $\Omega[n]\in\FS$ be the first $n$ bits of $\Omega$.

\begin{lmm}
\label{lmm:rec}
For $n=\bb^{-1}(m)$, $\K(\Omega[n]|m,n)=O(1)$.
\end{lmm}
\begin{prf}
For a string $x$, let $BB(x) = \inf\{t:\Omega^t>0.x\}$. Enumerate strings of length $n$, starting with $0^n$, and return the first string $x$ such that $BB(x)\geq m$. This string $x$ is equal to $\Omega[n]$, otherwise let $y$ be the largest common prefix of $x$ and $\Omega[n]$. Thus $BB(y)=\bb(\|y\|)\geq BB(x)\geq m$, which means $\bb^{-1}(m)\leq \|y\|<n$, causing a contradiction.\qed
\end{prf}

\begin{thr}
For $\{(x_i,y_i)\}_{i=1}^n$,  there exists $f:\N\rightarrow\N$ with $f(x_i)= y_i$ for $i\in\{1,\dots,n\}$ and $\K(f) \lel \sum_{i=1}^n\K(y_i|x_i)+\I(\{(x_i,y_i)\};\ch)$.
\end{thr}
\begin{prf}
Let $S=\{(x_i,y_i)\}$. Let $K=\sum_{i=1}^n\K(y_i|x_i)$.  We have $T=\arg\min_t\sum_{i=1}^n\K_t(y_i|x_i)=K$. Let $N=\bb^{-1}(T)$ and $M=\bb(N)$ and we define $m(x|y)=2^{-\K_M(x|y)}$, setting $m(\emptyset|y)=1-m(\N|y)$.

We condition all terms on $M$ and $K$, and later in the proof, we'll make this condition explicit. Let $Q$ be an elementary probability that realizes the stochasticity of $S$, where $d=\max\{\d(S|Q),1\}$. Without loss of generality, we can assume the support of $Q$ consists entirely of samples $R=\{(x_j,y_j)\}_{j=1}^{n_R}$ (of potentially different sizes) such that $\prod_{j=1}^{n_R}m(y_j|x_j)=2^{-M}$. Let 
$$z=\max\{ x : (x,y)\in R\in\mathrm{Support}(Q)\}.$$
We define a probability measure $\kappa$ over $d2^K$ lists $\mathcal{L}$ of size $z$ over $\N$, where each $\ell\in\mathcal{L}$ is chosen independently, and for each $\ell\in\mathcal{L}$, $\ell(i)$ is chosen independently according to $m(\cdot|i)$. We say a sample $R=\{(x_j,y_j)\}$ is inconsistent with a list $\ell$, $R\ltimes\ell$, if there exists $j$, where $\ell(x_j)\neq y_j$. $\eta(R,\mathcal{L})=[\forall\ell\in \mathcal{L}, E\ltimes \ell]$.
\begin{align*}
\E_{\mathcal{L}\sim\kappa}\E_{R\sim Q}[\eta(R,\mathcal{L})]=\E_{\{(x_j,y_j)\}\sim Q}\left(1-\prod_jm(y_j|x_j)\right)^{d2^K}<\E_{R\sim Q}e^{-d}=e^{-d}.
\end{align*}
Thus there exists a set of $d2^K$ lists $\mathcal{L}$, where $\E_{R\sim Q}[\eta(R,\mathcal{L})]<e^{-d}$. Thus let $t(R) = \eta(R,\mathcal{L})e^d$ be a $Q$-test, where $\E_{R\sim Q}[t(R)]\leq 1$. It must be $t(S)=0$, otherwise we have
$$
1.44d\leq \log t(S)\lea \d(S|Q) \lea d,
$$
which is contradiction for large $d$, which we can assume without loss of generality. So there exists a list $\ell$ such that $\ell(x_i)=y_i$, for all $(x_i,y_i)\in S$. Thus one can construct a total computable function $f:\N\rightarrow\N$ from $\ell$ that is consistent with $S$, for example $f(x)=\ell(x)$ if $x\leq z$ and $f(x)=1$ otherwise. Making the condition term $M$ explicit and keeping the condition term $K$ implicit we have,
\begin{align*}
\K(f|M) &\lea \K(\ell|M)\\
&\lea \log |\mathcal{L}| + \K(\mathcal{L}|M)\\
&\lea K +\log d+ \K(Q,d|M)\\
&\lea K +\Ks(S|M).
\end{align*}
Using Lemma \ref{lmm:ksh}, we get, noting $M=\bb(N)$, and $\bb$ is computable relative to $\ch$,
\begin{align*}
\K(f|M) &\lel K +\I(S;\ch\;|\;M).\\
\K(f) &\lel K +\K(S|M)+\K(M)-\K(S|\ch)+\K(N).
\end{align*}
So we have,
\begin{align}
\nonumber
&\K(S|M)+\K(M)\\
\nonumber
\lea& \K(S|M,\K(M))+\K(\K(M)|M)+\K(M)\\
\label{eq:chain}
\lea & \K(S,M)+\K(\K(M)|M)\\
\label{eq:NM}
\lea & \K(S,N,M)+O(\log N)\\
\label{eq:erM}
\lea & \K(S,N)+O(\log N).\\
\nonumber
\lea & \K(S)+O(\log N).\\
\label{eq1}
\K(f) \lel & K+ \K(S)-\K(S|\ch)+O(\log N).
\end{align}
Equation \ref{eq:chain} is from the chain rule. Equation \ref{eq:NM} is from the fact that $M=\bb(N)$. Equation \ref{eq:erM} comes $\K(T|S,K)=O(1)$ and Lemma \ref{lmm:rec}, which implies $\K(M|N,T) \lea \K(\Omega[N]|N,T) \lea  O(1)$. 

From $K$, and $S$, one can compute $T$, where $\bb^{-1}(T)=N$. Therefore by Lemma \ref{lmm:rec}, $\K(\Omega[N]|S)\lea \K(N)$, so by Lemma \ref{lmm:cons}, 
\begin{align}
\label{eq2}
N&\lel \I(\Omega[N];\ch)\lel \I(S;\ch)+\K(N)\lel \I(S;\ch). 
\end{align}
The above equation used the common fact that the first $n$ bits of $\Omega$ had $n-O(\log n)$ bits of mutual information with $\ch$. 
So combining Equations \ref{eq1} and \ref{eq2}, we get
$$
\K(f) \lel K+ \I(S;\ch).
$$
The proof is completed by noting the log precision, and the $K$ term in the equation removes the implicit conditioning of $K$.\qed
\end{prf}

\end{document}